\documentclass[11pt]{article}
\usepackage[a4paper,margin=28mm]{geometry}
\usepackage{amsmath,amsthm,amssymb}
\usepackage{booktabs,longtable,array,microtype,needspace}
\usepackage[numbers,sort&compress]{natbib}
\usepackage[hidelinks]{hyperref}
\hypersetup{pdftitle={Monodromy Computation Model},pdfauthor={Xiaoyang Liu},
  pdfsubject={A finite covering calculus and its computational characterization},
  pdfkeywords={monodromy, analytic continuation, computability, Lean}}
\numberwithin{equation}{section}
\newtheorem{theorem}{Theorem}[section]
\newtheorem{lemma}[theorem]{Lemma}
\newtheorem{proposition}[theorem]{Proposition}
\newtheorem{corollary}[theorem]{Corollary}
\theoremstyle{definition}
\newtheorem{definition}[theorem]{Definition}

\theoremstyle{remark}

\newcommand{\N}{\mathbb N}
\newcommand{\Z}{\mathbb Z}
\newcommand{\R}{\mathbb R}
\newcommand{\C}{\mathbb C}
\newcommand{\Sone}{\mathbb S^1}
\newcommand{\Part}{\operatorname{Part}}

\newcommand{\ev}{\operatorname{ev}}
\newcommand{\Sel}{\operatorname{sel}}
\newcommand{\Gate}{\operatorname{gate}}

\newcommand{\bin}{\operatorname{bin}}
\newcommand{\TM}{\operatorname{TM}}
\newcommand{\code}[1]{\nolinkurl{#1}}
\newcommand{\step}{\operatorname{step}}

\DeclareMathOperator{\Rea}{Re}
\title{Monodromy Computation Model}
\author{Xiaoyang Liu\thanks{Source code:
  \url{https://github.com/ERHUTUZI123/monodromy}.}\\[0.2em]
  \small University of Waterloo\\
  \small\href{mailto:x98liu@uwaterloo.ca}{x98liu@uwaterloo.ca}}
\date{November 2026}

\begin{document}
\maketitle

\begin{abstract}
A closed path can return to its starting point while analytic continuation
reaches another branch. We construct a finite computational calculus in
which retained logarithmic branches supply memory, analytic observations
select operations, and descent through a covering exposes deeper memory.
Three branch coordinates suffice, and every finite observation uses the
same cubic power covering. Each program defines a globally real analytic
return map whose integer-lattice execution is realized by actual path
lifting. Our first theorem simulates every finite-alphabet, finite-control
Post--Turing machine, preserving complete configurations, halting, and
divergence. Our second theorem identifies the partial numeric functions
of the calculus, under fixed primitive-recursive input and output maps,
with exactly the partial recursive functions. Deleting descent leaves a
Forward language with a syntax-derived finite observation quotient and a
primitive-recursive halting predicate. This yields a strict separation
within the calculus and identifies access to deeper memory as its decisive
additional operation. The constructions and semantic correspondences are
formalized in Lean~4 with Mathlib.
\end{abstract}

\section{Introduction}

Digital computation is described by discrete state transitions.
Continuous-time models such as Shannon's general-purpose analog
computer~\cite{shannon1941} and real-arithmetic models such as the
Blum--Shub--Smale machine~\cite{blumShubSmale1989} organize computation
through different state spaces and primitive operations. Here the
organizing structure is branch transport under analytic continuation.

A closed path can change an analytic branch. Continuing a logarithm
around the origin changes its value by $2\pi i$ times the winding
number. The retained branch records this integer, and endpoint-fixed
homotopies preserve its transport~\cite[\S1.3]{hatcher2002}.
Can this geometry support memory, conditional execution, and universal
computation within a finite effective calculus?

We introduce the \emph{Monodromy computation calculus}. Logarithmic
branches retain integers; a cubic covering observes their residues;
analytic selectors turn those observations into internal control.
Affine writing and covering-level descent expose successive digits
of memory. Finite expressions determine each return and its stopping
signal, and generate paths whose actual lifts reach the next state.
The contributions are:
\begin{enumerate}
\item A finite calculus with three retained branch coordinates, integer
affine updates, analytic selection, descent, and internal feedback.
Each program has a globally real analytic return map and a path-lifting
realization of every finite run.
\item \textbf{Theorem I:} a native simulation of every finite-alphabet,
finite-control Post--Turing machine, preserving each primitive step,
the complete configuration, halting, and divergence.
\item \textbf{Theorem II:} under fixed primitive-recursive input and
output maps, $\mathcal C_{\mathrm{Mono}}=\mathcal R_{\mathrm{partial}}$.
Deleting descent gives a Forward language with primitive-recursive
halting. An undecidable partial-recursive domain proves strict
separation between the two languages.
\item A Lean~4 formalization of the definitions, compilation,
complete-execution correspondences, and structural separation.
\end{enumerate}

Analytic simulation of Turing machines has a substantial history.
Koiran and Moore~\cite{koiranMoore1999} use integer encodings, analytic
selectors, and radix division to obtain universal analytic maps.
Gra\c ca, Campagnolo, and Buescu~\cite{gracaCampagnoloBuescu2005}
develop robust analytic simulations, and Gra\c ca and
Zhong~\cite{gracaZhong2023} obtain robust one-dimensional analytic maps.
Our contribution organizes these established computational ingredients
through a finite covering syntax, actual transport semantics, a fixed
effective interface, and a descent deletion theorem. The latter
identifies a structural difference between retaining unbounded state
and being able to access it in future computation.

\paragraph{Conventions.}
We write $[a]=\{0,\ldots,a-1\}$. Division of integers by a positive
integer is Euclidean: $k=B(k\div B)+(k\bmod B)$, with
$0\leq k\bmod B<B$, also when $k<0$.
A value of $\Part A$ is either a result in $A$ or undefined. For a
partial value $v$, the expression $f_*v$ means its image under a total map
$f$; it is undefined exactly when $v$ is undefined. Thus equalities in
$\Part A$ preserve both results and divergence.

\section{Mathematical Preliminaries}
\label{sec:foundations}

\subsection{Transport and logarithmic memory}

A covering supplies a precise meaning for carrying a branch. A map
$p:E\to X$ is a covering when every point of $X$ has an open
neighborhood whose preimage is a disjoint union of open sets, each
mapped homeomorphically onto that neighborhood. For a path
$\gamma:[0,1]\to X$ and
each $e\in p^{-1}(\gamma(0))$, there is a unique lift starting at $e$.
Its endpoint defines transport
\[
 \mathcal T_\gamma:p^{-1}(\gamma(0))\longrightarrow p^{-1}(\gamma(1)).
\]
If $\gamma*\delta$ means first $\gamma$, then $\delta$, the standard
lifting theorems give~\cite[\S1.3]{hatcher2002}
\begin{equation}\label{eq:transport}
 \mathcal T_{\gamma*\delta}=\mathcal T_\delta\circ\mathcal T_\gamma,
 \qquad \mathcal T_{\bar\gamma}=\mathcal T_\gamma^{-1}.
\end{equation}
Endpoint-fixed homotopies preserve transport. For loops, these endpoint
maps form the monodromy action on a fiber.

For the complex exponential $\exp:\C\to\C^*$ and the power maps on
$\C^*$, local inverse branches are holomorphic. Carrying such a
branch along a path is analytic continuation, represented by the path's
lift. A Riemann surface separates different branches into distinct
points over a common base~\cite{forster1981}. The construction below
uses the explicit exponential and power coverings, together with their
regular-root selectors.

For memory we use the exponential's real angular restriction
\begin{equation}\label{eq:log-cover}
 p:\R\longrightarrow\Sone,\qquad p(\theta)=e^{i\theta}.
\end{equation}
Its fiber above $1$ is $2\pi\Z$. These are the angular representatives
of logarithmic branches. We use the normalized coordinate $k\in\Z$ for
the retained angle $2\pi k$. The loop
$t\mapsto e^{2\pi i nt}$ transports $k$ to $k+n$: its lift is
$t\mapsto2\pi(k+nt)$. In particular, a fixed loop acts by the same
translation on every branch.

Retaining a point of the fiber is essential. The common base value $1$
alone contains no record of the integer $k$. Several retained branches
will be represented by a vector $x\in\Z^n$.

\subsection{A finite observation of an infinite branch}

To inspect one digit of a retained integer, factor the exponential
covering through a power covering. For $B\geq1$, put
\begin{equation}\label{eq:root-observation}
 r_B(x)=e^{2\pi i x/B},\qquad P_B(z)=z^B\quad(z\in\C^*).
 \qquad P_B(r_B(x))=e^{2\pi i x}.
\end{equation}
For integer $x$, the observed root $r_B(x)$ depends exactly on $x\bmod B$.
The state retains all of $x$, while this observation has only $B$ values.
Throughout the computational calculus we fix $B=3$.

This observation is itself compatible with path lifting. For a continuous
real curve $u(t)$, $r_B(u(t))$ is the lift through $P_B$ of
$e^{2\pi i u(t)}$ starting at $r_B(u(0))$. Uniqueness of lifting proves
that its endpoint is $r_B(u(1))$. Each expression using an observation
will have its own induced curve of this form.

\subsection{Regular roots as analytic selectors}

A finite observation becomes a choice of operation through an analytic
interpolant. Suppose $f:\C\to\C$ is entire and $r$ is a regular point,
so $f'(r)\ne0$. Define the normalized divided-difference kernel
\begin{equation}\label{eq:kernel}
 K_r^f(z)=\frac{f(z)-f(r)}{f'(r)(z-r)},\qquad K_r^f(r)=1.
\end{equation}
The singularity at $r$ is removable. Consequently $K_r^f$ is entire;
on the fiber $f^{-1}(f(r))$ it equals $1$ at $r$ and $0$ at every other
point. If $r_0,\ldots,r_{B-1}$ are distinct regular roots in one fiber,
then $\sum_d v_d K_{r_d}^f(z)$ selects $v_j$ at $z=r_j$.

For the cubic cover, let $\zeta_d=e^{2\pi i d/3}$, $d\in[3]$.
The kernels and their real observations are
\begin{equation}\label{eq:cubic-weights}
 K_d(z)=\frac{z^2+\zeta_d z+\zeta_d^2}{3\zeta_d^2},
 \qquad W_d(x)=\Rea K_d(r_3(x)).
\end{equation}
The polynomial formula includes the removable value in
\eqref{eq:kernel}. Each $W_d$ is real analytic on $\R$, and
\begin{equation}\label{eq:kronecker}
 W_d(k)=\begin{cases}1&k\bmod3=d,\\0&k\bmod3\ne d\end{cases}
 \qquad(k\in\Z).
\end{equation}
Thus a finite choice is an analytic formula, evaluated at the current
retained state.

The kernels also respect hierarchical composition. For entire $f,g$
with $f'(r)\ne0$ and $g'(f(r))\ne0$, the divided-difference chain rule gives
\begin{equation}\label{eq:kernel-composition}
 K_r^{g\circ f}(z)=K_{f(r)}^g(f(z))K_r^f(z).
\end{equation}
When $f(z)\ne f(r)$ the factors telescope. When $z\ne r$ and
$f(z)=f(r)$ both sides vanish, and at $z=r$ both sides equal one.
This is a normalized instance of the classical chain
rule for divided differences~\cite{floaterLyche2007}. It explains how
selection at one level can be refined by selection at the next. Our
finite syntax will implement the hierarchy by nested cubic queries.

\section{Syntax and Semantics of the Monodromy Calculus}
\label{sec:model}

The universal calculus retains three normalized logarithmic coordinates
$x=(x_0,x_1,x_2)\in\Z^3$, representing the angles $2\pi x_j$ above
the common base point $1$. In the Turing realization, $x_0$ stores
finite control and the current symbol, and $x_1,x_2$ store two words.
We first describe the memory operations, then give their finite syntax
for an arbitrary finite number $n$ of coordinates. Both main theorems
use $n=3$.

\subsection{Memory, writing, and descent}

The selectors provide both a branch query and its companion descent:
\begin{equation}\label{eq:select-descent}
 S(x;y_0,y_1,y_2)=\sum_{d=0}^2W_d(x)y_d,
 \qquad
 D(x)=\frac{x-\sum_{d=0}^2 dW_d(x)}3.
\end{equation}
For integers, these formulas satisfy
\begin{equation}\label{eq:select-descent-int}
 S(k;y_0,y_1,y_2)=y_{k\bmod3},\qquad D(k)=k\div3.
\end{equation}
The second equality is the Euclidean decomposition of $k$. Descent
removes a digit whose value has first been observed; applying the same
observation to the result reveals the following digit.

For binary words, reserve zero for the empty word and use the nonzero
ternary digits for letters:
\begin{equation}\label{eq:binary-code}
 E([])=0,\qquad E(b::w)=b+1+3E(w),\qquad b\in\{0,1\}.
\end{equation}
This is injective. Indeed, the residue modulo three distinguishes an
empty word from its first letter, and division by three recovers the
tail; induction recovers the whole word. Prepending $b$ is the affine
write $x\mapsto b+1+3x$, and $D$ removes the first letter.
For example,
\[
 E([0,1])=7,\qquad 7\xrightarrow{\text{read }1}D(7)=2
 \xrightarrow{\text{read }2}D(2)=0.
\]
The two observations identify the letters $0$ and $1$, followed by the
empty word. The same cubic covering is used throughout.

\subsection{Finite expressions and guards}

A program must give a finite description of every operation it can
perform. Fix $n\geq1$ coordinates. Its expressions are generated by
\begin{equation}\label{eq:grammar}
 t ::= x_j\mid c\mid t+t\mid c\,t
       \mid\Sel(t;t_0,t_1,t_2)\mid D(t),
 \qquad j\in[n],\quad c\in\Z.
\end{equation}
The integer interpretation $\ev_\Z(t,x)$ uses ordinary integer
arithmetic, selects branch $\ev_\Z(t,x)\bmod3$, and interprets $D$
as Euclidean division by three. The real interpretation
$\ev_\R(t,u)$ uses the same arithmetic and the formulas
\begin{align}
 \ev_\R(\Sel(t;t_0,t_1,t_2),u)
   &=\sum_{d=0}^2W_d(\ev_\R(t,u))\ev_\R(t_d,u),\label{eq:real-select}\\
 \ev_\R(D(t),u)&=D(\ev_\R(t,u)).\label{eq:real-descend}
\end{align}
Both the key and the possible outcomes can depend on the whole current
state. All expressions are finite trees.

Stopping is specified in the same language of finite observations. A
guard is generated by
\[
 h ::= \mathsf{false}\mid\mathsf{true}
       \mid\Sel(t;h_0,h_1,h_2).
\]
Its integer interpretation is Boolean. Replacing its constant leaves by
$0$ and $1$ gives an expression $\widehat h$ and hence a real analytic
indicator $H(u)=\ev_\R(\widehat h,u)$. For a guard $h$ and expressions
$a,b$, define $\Gate(h;a,b)$ by replacing each true leaf of $h$ by
$a$ and each false leaf by $b$, preserving all query nodes. This is
again an expression of~\eqref{eq:grammar}.

\begin{lemma}[Exact analytic interpretation]\label{lem:interpretation}
Every expression defines a real analytic function on $\R^n$, and
\[
 \ev_\R(t,x)=\ev_\Z(t,x)\qquad(x\in\Z^n),
\]
where integers on the right are embedded in $\R$. A guard indicator
is $1$ exactly at its true integer states and $0$ at its false integer
states. Moreover,
\[
 \ev_\Z(\Gate(h;a,b),x)=
 \begin{cases}\ev_\Z(a,x)&h(x)=\mathsf{true},\\
               \ev_\Z(b,x)&h(x)=\mathsf{false}.
 \end{cases}
\]
\end{lemma}
\begin{proof}
Use simultaneous structural induction. Coordinates and constants are
analytic, and sums and products preserve analyticity. A query is the
finite sum~\eqref{eq:real-select} of analytic compositions and products;
descent is~\eqref{eq:real-descend}. On integer states, each induction
hypothesis makes the key an integer, so~\eqref{eq:select-descent-int}
gives the required query or quotient. The guard and gate assertions
follow by induction on the guard tree.
\end{proof}

\subsection{Returns, stopping, and complete execution}

The program consists of a guard and an update expression for each
coordinate.
\begin{definition}[Program and execution]\label{def:program}
A program $P=(h,u_0,\ldots,u_{n-1})$ has returning expressions
$r_j=\Gate(h;x_j,u_j)$. Write
\begin{equation}\label{eq:return}
 T_P(x)_j=\ev_\Z(r_j,x),\qquad
 F_P(v)_j=\ev_\R(r_j,v).
\end{equation}
Its partial step stops at $x$ when $h(x)$ is true, and otherwise continues
at $T_P(x)$. The complete result $\mathcal E_P(x)\in\Part\Z^n$ is
$T_P^k(x)$ at the first $k\geq0$ with $h(T_P^k(x))$ true, and is
undefined if no such $k$ exists.
\end{definition}

By Lemma~\ref{lem:interpretation}, $F_P$ is globally jointly real
analytic, preserves the integer lattice, and restricts there to $T_P$.
At a stopped integer state $T_P(x)=x$, so returns can be counted after
halting by keeping the state stationary. The native memory domain is
the exact lattice $\Z^n$ inside $\R^n$, with its real coordinates and
retained exponential branches. Its native step tests the internally
generated indicator $H=1$ and otherwise applies $F_P$.

Let $\iota$ be the inclusion of an integer state into this native domain.
Induction on each step and the first stopping time yields the full
partial-value equality
\begin{equation}\label{eq:native-eval}
 \mathcal E_P^{\mathrm{native}}(\iota x)=\iota_*\mathcal E_P(x).
\end{equation}
The same induction shows that an infinite integer run has an infinite
native run. Conversely, every native step on this domain is the image of
the integer step. Thus~\eqref{eq:native-eval} also preserves divergence.

The guard, branch choices, and writes are functions of the current
state and the fixed finite syntax. That syntax is the runtime
controller; a source Turing table enters when the program is constructed.
Execution consists of discrete returns, with the stopping signal read
at those returns. The following path generator realizes each return.

\subsection{Actual continuation paths}

We now connect a return to transport rather than merely identifying its
endpoint symbolically. For a native state $x\in\Z^n$, define the
coordinate angle curves and their base loops by
\begin{equation}\label{eq:generated-path}
 \ell_{P,x,j}(t)=2\pi\big((1-t)x_j+tF_P(x)_j\big),\qquad
 \gamma_{P,x,j}(t)=e^{i\ell_{P,x,j}(t)}.
\end{equation}
Both endpoints of $\gamma_{P,x,j}$ are $1$. The retained starting
angle is $2\pi x_j$. The curve $\ell_{P,x,j}$ is a lift of the
generated loop with this starting point, so uniqueness of lifting gives
\begin{equation}\label{eq:step-lift}
 \mathcal T_{\gamma_{P,x,j}}(2\pi x_j)=2\pi T_P(x)_j.
\end{equation}
The program's internal analytic rules first determine its return; the
interpolation in~\eqref{eq:generated-path} then generates a path
realizing that return through the exponential covering.

\begin{proposition}[Finite-run lifting]\label{prop:lifting}
Concatenate the generated loops for $k$ consecutive returns, updating
the state after each return. The resulting path $\Gamma_{P,k,x,j}$
satisfies
\begin{equation}\label{eq:run-lift}
 \mathcal T_{\Gamma_{P,k,x,j}}(2\pi x_j)=2\pi(T_P^k(x))_j.
\end{equation}
Every query node also follows actual transport through the cubic power
covering along its own induced base path.
\end{proposition}
\begin{proof}
Equation~\eqref{eq:step-lift} proves the case of one return, and
transport composition~\eqref{eq:transport} proves the finite-run
statement by induction. For a query key $t$ along a continuous state
curve $v(s)$, use the continuous real curve
$u(s)=\ev_\R(t,v(s))$. Equation~\eqref{eq:root-observation} identifies
$r_3(u(s))$ as the unique lifted query curve. Its endpoint at an integer
state is the root indexed by $\ev_\Z(t,x)\bmod3$.
\end{proof}

Endpoint-fixed homotopies of the generated base paths preserve these
transport results. This statement concerns the path realization of a
specified return. Its computational meaning comes from the finite
analytic rule that generated the path.

\section{Theorem I: Turing Universality}
\label{sec:universality}

\subsection{Arbitrary finite tables from cubic queries}

Repeated observation and descent read an arbitrary finite ternary
address. Let $Q_d(t;A)$ be the expression querying a table $A$ with
$3^d$ entries. Define
\begin{align}
 Q_0(t;A)&=A_0,\nonumber\\
 Q_{d+1}(t;A)
   &=\Sel\bigl(t;
       Q_d(D(t);(A_{0+3j})_j),
       Q_d(D(t);(A_{1+3j})_j),
       Q_d(D(t);(A_{2+3j})_j)\bigr).
 \label{eq:query-tree}
\end{align}
Induction using $k=(k\bmod3)+3(k\div3)$ proves
\begin{equation}\label{eq:query-tree-exact}
 \ev_\Z(Q_d(t;A),x)
   =\ev_\Z(A_{\ev_\Z(t,x)\bmod3^d},x).
\end{equation}
The same construction works for guards. Every tree is finite, and every
node uses the same cubic query or descent.

For an alphabet $[a]$ with $a\geq1$, encode letter $j$ as the binary
word of length $a$ whose only $1$ is at position $j$. Its ternary block
value is
\begin{equation}\label{eq:block-code}
 \delta_j=\sum_{i=0}^{a-1}(1+\mathbf1_{i=j})3^i,
 \quad B=3^a,\qquad
 E_a([])=0,\quad E_a(j::w)=\delta_j+B E_a(w).
\end{equation}
The $\delta_j$ are distinct and satisfy $0<\delta_j<B$. A query of
depth $a$ therefore reads the first letter, including the empty-word
sentinel, and $a$ descents expose the tail. The recurrence proves
injectivity of $E_a$ by the same residue-and-quotient argument as for
\eqref{eq:binary-code}. This block encoding accommodates any finite
alphabet while keeping the covering degree fixed.

\subsection{Two stacks and a finite controller}

We use the Post--Turing convention formalized as Mathlib's
$\TM_0$~\cite{mathlibComputability}. A table has a finite alphabet
$\Sigma$ with a distinguished blank and a finite control set $Q$ with
an initial label. Each pair $(q,s)$ either halts or specifies a new
label together with one primitive instruction: write a symbol, move
left, or move right. A conventional write-and-move instruction is
represented by a finite sequence of these primitives.

A complete configuration can be represented by
\[
 c=(q,L,s,R),
\]
where $L$ and $R$ are finite words, nearest symbols first, and $s$ is
the current symbol. Words represent blank-extended half-tapes:
appending trailing blanks gives the same half-tape. Every finitely
supported tape has such a representative. Let $\operatorname{head}_0$
return the first symbol, or the blank on an empty word. A right move
with target label $q'$ is
\begin{equation}\label{eq:stack-move}
 (q,L,s,R)\longmapsto
 (q',s::L,\operatorname{head}_0(R),\operatorname{tail}(R)).
\end{equation}
A left move is symmetric, and writing changes only $s$ and the label.
These operations induce exactly the usual operations on the entire
blank-extended tape.

Choose blank-preserving enumerations $\Sigma\simeq[a]$ and $Q\simeq[q]$,
with the blank and initial label numbered zero. Encode a stack
configuration by three integers:
\begin{equation}\label{eq:configuration-code}
 C(q_0,L,s,R)=\bigl(aq_0+s,\ E_a(L),\ E_a(R)\bigr).
\end{equation}
The first coordinate jointly stores the finite controller and the
current symbol; the other coordinates store the two unbounded words.

The integer return row for a right move is explicit. If
$x=(c,l,r)$ encodes $(q_0,L,s,R)$, set
\begin{equation}\label{eq:right-row}
 (c,l,r)\longmapsto
 \bigl(aq'+\operatorname{head}_0(r),\ \delta_s+Bl,\ r\div B\bigr),
\end{equation}
where $\operatorname{head}_0(r)$ is a finite table on $r\bmod B$,
returning the decoded letter at a valid block and blank at zero.
Its other entries can be assigned arbitrarily, for instance blank.
Equation~\eqref{eq:query-tree-exact} implements this table. A left
row exchanges the stack roles, and a write row is
$(aq'+s',l,r)$. Thus every row belongs to the finite cubic syntax.

There are $aq$ possible control-symbol pairs. A table query of depth
$aq$ selects the corresponding row from coordinate $c$; indeed
$aq<3^{aq}$. Addresses outside the source table are stationary halting
entries. A guard tree with the same address reports exactly the halting
entries of the source. The compiled program is denoted $P_M$.

\begin{lemma}[One-step compilation]\label{lem:one-step}
For every stack configuration $c$, including halting ones,
\begin{equation}\label{eq:compiler-step}
 \step_{P_M}(C(c))=C_*\step_M(c).
\end{equation}
Here both sides are optional successors: a halting step has no successor.
The code $C$ is injective on stack configurations.
\end{lemma}
\begin{proof}
The control query selects the entry indexed by $aq_0+s$.
For a write, the first coordinate changes to the prescribed label and
symbol and both stacks stay fixed. For a right move,
$E_a(s::L)=\delta_s+BE_a(L)$, while
$E_a(R)\div B=E_a(\operatorname{tail}R)$ and its block residue
recovers $\operatorname{head}_0(R)$. These are exactly
\eqref{eq:stack-move} and~\eqref{eq:right-row}. The left case is
symmetric. The compiled guard is true exactly when the source entry
halts. Finally, the first code coordinate recovers $(q_0,s)$ by finite
product enumeration and the other two recover their words by
injectivity of $E_a$.
\end{proof}

\Needspace{12\baselineskip}
\subsection{The full simulation theorem}

\begin{theorem}[Theorem I: Turing simulation]\label{thm:universality}
For every finite-alphabet, finite-control Post--Turing machine $M$,
there is a program $P_M$ with three retained logarithmic branches and
only cubic queries and descents, with the following properties.
Every complete source configuration has a representing native state.
From every such representation, one source primitive step is one native
return. For each number of returns, the native state represents the
entire source configuration, with halting configurations kept stationary.
Complete execution preserves the final configuration, halting, and
divergence in both directions. Every finite native run is realized by
the actual transports in~\eqref{eq:run-lift}, and its return map is
globally jointly real analytic.
\end{theorem}
\begin{proof}
First enumerate the finite alphabet and control, preserving their
distinguished elements. A complete source tape has representatives
$L,R$ as above. Initialize the native state with $C(q_0,L,s,R)$.
Lemma~\ref{lem:one-step} and induction give the exact stack execution
at every return. Passing to blank-extended half-tapes commutes with
each move and write, so this execution represents the whole source
tape, independently of trailing-blank representatives.

At any represented configuration, the source has no successor exactly
when the native guard is true. Consequently, a source halt yields a
native halt at the same return with the represented final
configuration. Conversely, the first native halt is a source halt,
since its guard is precisely the source table's stopping predicate.
If either run continues forever, this equivalence at every finite
return forces the other to continue forever. Equivalently, on stack
representatives the complete partial values satisfy
\begin{equation}\label{eq:compiled-part}
 \mathcal E_{P_M}(C(c))=C_*\mathcal E_M^{\mathrm{stack}}(c),
\end{equation}
and decoding the right side gives the complete source result.
Equation~\eqref{eq:native-eval} transfers this equality to native
execution. Lemma~\ref{lem:interpretation} gives global analyticity,
and Proposition~\ref{prop:lifting} supplies each finite run's actual
lifted paths.
\end{proof}

\subsection{A complete computation}

Consider a two-symbol alphabet $[2]$ with blank $0$, two controls
$q_0=0,q_1=1$, and the table
\[
 (q_0,0)\mapsto(q_1,\text{write }1),\qquad
 (q_1,1)\mapsto(q_0,\text{move right}),
\]
with all other entries halting. Start with current symbol $0$ and
right word $[1]$. For $a=2$, the blocks of letters $0$ and $1$ are
$[1,0]$ and $[0,1]$, so $\delta_0=5$, $\delta_1=7$, and $B=9$.
The entire run is
\begin{center}
\begin{tabular}{@{}cclc@{}}
\toprule
Return & Source configuration & Integer state & Action\\
\midrule
$0$ & $(q_0,[],0,[1])$ & $(0,0,7)$ & write $1$\\
$1$ & $(q_1,[],1,[1])$ & $(3,0,7)$ & move right\\
$2$ & $(q_0,[1],1,[])$ & $(1,7,0)$ & halt\\
\bottomrule
\end{tabular}
\end{center}
The first return changes the control coordinate by $3$. The second
changes the three coordinates by $(-2,7,-7)$. Thus their exponential
base loops have these respective integer windings; lifting from the
retained branches gives exactly the displayed states. The halting
guard is generated by the same finite control query as the updates.
This example is a direct instance of the compilation rules; the
general theorem proves the same correspondence for arbitrary finite
tables and unbounded word lengths.

\section{Theorem II: Characterization and Sublanguage Separation}
\label{sec:characterization}

\subsection{A fixed numeric interface}

A computational characterization must account for the encoding of
both inputs and results. We fix once and for all two total maps
\begin{equation}\label{eq:io}
 I:\N\longrightarrow\Z^3,\qquad O:\Z^3\longrightarrow\N.
\end{equation}
They are independent of the requested function and the program.
The input map prepares a canonical binary numeric tape over the fixed
alphabet used by Mathlib's partial-recursive-to-Turing translation.
It then stores the current symbol, an empty left word, and the encoded
right word in the three coordinates. The output map decodes only the
right word, reads its main-track bits in order, and ignores separators
and blank symbols. Appendix~\ref{app:io} specifies both maps, including
their values on malformed integer states.

\begin{lemma}[Effective interface]\label{lem:io}
The fixed maps $I$ and $O$ are primitive recursive. If a completed
configuration is the canonical numeric result for $n$, then its native
code has output $n$.
\end{lemma}
\begin{proof}
Finite-symbol encoding, binary numeral construction, list operations,
and the recurrence~\eqref{eq:block-code} are primitive recursive.
For decoding a natural code $v$, scan its lowest block and replace
$v$ by $v\div3^a$. A valid nonempty block code decreases, and the
encoded number bounds the word length. Running exactly the original
$v$ scan steps, with zero or an invalid block ending the scan, gives
a total primitive-recursive decoder. On valid codes it inverts $E_a$.
Clamp negative integer coordinates to zero. The output fold replaces
an accumulator $z$ by $2z+b$ at a bit $b$ and leaves it unchanged at
every other symbol. It is primitive recursive and reads the canonical
numeric tape as $n$. Trailing blanks leave the fold unchanged, so
the result also holds for any representative of the same half-tape.
\end{proof}

\begin{definition}[Realized numeric function]\label{def:realizable}
A program $P$ computes $f:\N\to\Part\N$ when, for every $n$,
\begin{equation}\label{eq:computes}
 O_*\mathcal E_P(I(n))=f(n).
\end{equation}
Let $\mathcal C_{\mathrm{Mono}}$ consist of the functions so computed
by three-coordinate programs of~\eqref{eq:grammar}. Equivalently,
by~\eqref{eq:native-eval}, the native execution has a result representing
an output $m$ exactly when $m$ is the result of $f(n)$.
\end{definition}

\subsection{Upper bound: effective syntax}

\begin{lemma}[Effectiveness of finite execution]\label{lem:effective}
For each fixed finite program $P$, its integer next-state map and
guard are primitive recursive. Its complete execution $\mathcal E_P$
is partial recursive under a standard finite-vector integer encoding.
\end{lemma}
\begin{proof}
Integer addition, multiplication by a fixed integer, and Euclidean
division and remainder by a positive fixed integer are primitive
recursive in the signed integer encoding. Structural induction on
expressions therefore proves the claim for their interpretations:
a query is a case distinction over three primitive-recursive branch
values, and descent is division by three. Induction on guards proves
their claim, including the gate used for the returning expressions.
The partial step, with its optional successor, is consequently
primitive recursive.

Iterate this effective step until it stops. Formally, the loop body
returns either the current state as a final result or a successor to
iterate. Unbounded iteration of such an effective body is partial
recursive. Alternatively, search for the least $k$ with the
primitive-recursive predicate $h(T_P^k(x))$ true, and return
$T_P^k(x)$. This computes exactly Definition~\ref{def:program}, with
the search undefined precisely on the infinite runs.
\end{proof}

\begin{proposition}[Computational upper bound]\label{prop:upper}
Every function in $\mathcal C_{\mathrm{Mono}}$ is partial recursive.
\end{proposition}
\begin{proof}
Equation~\eqref{eq:computes} expresses the function as the composition
of the primitive-recursive input, the partial-recursive execution,
and the primitive-recursive output. These classes are closed under
the indicated composition.
\end{proof}

\subsection{Lower bound: preserving numeric results}

For completeness, we describe the classical computation-theoretic
bridge used in addition to Theorem~\ref{thm:universality}. The
partial recursive functions are generated from zero, successor, and
projections by composition, primitive recursion, and unbounded search.
Each has a finite code and can be executed by a finite-control stack
machine on a fixed finite alphabet. Composition stores intermediate
results; primitive recursion keeps a counter and accumulator; unbounded
search increases a candidate until its defining test returns zero,
continuing forever when that computation supplies no result.
These constructions are the standard route from partial recursion to
Turing computation. Mathlib supplies their formal implementation and
the translations from stacks to a tape and then to
Post--Turing primitives~\cite{carneiro2019,mathlibComputability}.

The following form records the exact interface needed here. For every
partial recursive $f$, there is a source code $c$ and a finite-support
Post--Turing table $M_c$ over one fixed alphabet such that
\begin{equation}\label{eq:classical-compiler}
 \mathcal E_{M_c}(\operatorname{init}(W_n))
       =(\operatorname{halt}_c\circ[\,\cdot\,])_*f(n).
\end{equation}
Here $W_n$ is the fixed numeric input word, and
$\operatorname{halt}_c([m])$ is the complete canonical final
configuration carrying numeric result $m$. It includes the final
label and the entire tape. The translated code visits only finitely
many labels: induction over its finite syntax supplies a closed finite
support containing the initial label. Restricting to that support
makes the control set finite while preserving every reachable step.

The complete partial equality~\eqref{eq:classical-compiler} follows
through the translations, rather than just a stopping equivalence.
The stack-to-tape representation determines the whole packed tape
from all its track projections. Hence each terminal stack configuration
has its specified canonical terminal tape; the later primitive
translation preserves that complete configuration. This also gives
the converse implication for termination and preserves divergence.

\begin{proposition}[Computational lower bound]\label{prop:lower}
Every partial recursive $f:\N\to\Part\N$ belongs to
$\mathcal C_{\mathrm{Mono}}$.
\end{proposition}
\begin{proof}
Choose $M_c$ as in~\eqref{eq:classical-compiler}, restrict to its
closed finite support, enumerate its control with the initial label
zero, and apply the cubic compilation of Section~\ref{sec:universality}.
The source alphabet and $W_n$ are fixed independently of $c$.
Since the initial label is zero, the code of the initial stack
configuration is exactly $I(n)$: its first coordinate is just the
numbered current symbol, regardless of the number of control labels.
The remaining coordinates are zero and the encoded right word.

The exact partial equalities of the source translations and
\eqref{eq:compiled-part} identify every native terminal state with
the corresponding canonical result tape. Lemma~\ref{lem:io} reads
that state as $m$. Thus if $f(n)$ returns $m$, native execution returns
a state read as $m$; if native execution returns a state, the source
has returned, and its unique numeric result is its readout. If $f(n)$
is undefined, the complete-execution equalities force native execution
to be undefined as well. This is precisely~\eqref{eq:computes}.
\end{proof}

\begin{theorem}[Theorem II: exact characterization]\label{thm:characterization}
Under the fixed primitive-recursive maps $I$ and $O$,
\begin{equation}\label{eq:characterization}
 \mathcal C_{\mathrm{Mono}}=\mathcal R_{\mathrm{partial}},
\end{equation}
where $\mathcal R_{\mathrm{partial}}$ is the class of partial recursive
functions $\N\to\Part\N$. Each realizing program uses three retained
branches and the same degree-three power covering for every finite
query and descent. Its native return map is globally jointly real
analytic and its finite runs satisfy~\eqref{eq:run-lift}.
\end{theorem}
\begin{proof}
Propositions~\ref{prop:upper} and~\ref{prop:lower} give the two
inclusions. Lemma~\ref{lem:interpretation} and
Proposition~\ref{prop:lifting} supply the analytic and transport
properties for the very same realizing program.
\end{proof}

\subsection{Forward execution and its finite observation}

The characterization establishes the capacity of the whole calculus.
To isolate the contribution of descent, call a term, guard, or program
\emph{Forward} when every expression in it is built without $D$.
The remaining operations still allow unbounded integer growth and
arbitrarily nested finite queries.

\begin{lemma}[Closed residue observation]\label{lem:forward-residue}
For each Forward program $P$ in $n$ coordinates, one can compute from
its syntax a positive integer $M$ such that coordinatewise congruence
modulo $M$ is preserved by $T_P$ and determines the guard $h$.
\end{lemma}
\begin{proof}
More generally allow any finite query degree $B\geq1$ in the syntax;
the cubic language is its special case. Define a syntactic support
certificate at $M$ recursively. Coordinates and constants are supported;
sums and integer scaling preserve support of their subterms. A query
of degree $B$ is supported when $B\mid M$ and its key and all branches
are supported at $M$. Induction on a certificate proves that the
term's value modulo $M$ depends only on the input modulo $M$: at a
query, congruent keys have equal residues modulo $B$, select the same
branch, and give congruent results.

A positive sufficient modulus is obtained recursively: use $1$ for
coordinates and constants, multiply moduli at sums, preserve the
modulus at integer scaling, and at a query multiply $B$, the key's
modulus, and all branch moduli. Induction on support certificates
shows that they persist on replacing a modulus by a multiple.
The analogous recursion on guard trees makes
their Boolean values equal on congruent inputs. Take a common product
for the guard and all updates. The guarded return then preserves
congruence, since congruent states make the same stopping choice.
All these are finite computations on syntax.
\end{proof}

Consequently the observation $\pi_M(x)=(x_j\bmod M)_j$ evolves
autonomously on the finite set $(\Z/M\Z)^n$. It carries a well-defined
successor $\overline T_P$ and stopping predicate $\overline h$:
\[
 \pi_M T_P=\overline T_P\pi_M,\qquad
 h=\overline h\pi_M.
\]
This quotient preserves the whole observed partial execution, including
its domain of definition.

\begin{theorem}[Forward halting and strict separation]\label{thm:forward}
For a Forward program $P$ and a syntax-derived modulus $M>0$,
\begin{equation}\label{eq:forward-bound}
 \mathcal E_P(x)\text{ is defined}
 \quad\Longleftrightarrow\quad
 \exists k<M^n\; h(T_P^k(x))=\mathsf{true}.
\end{equation}
Its halting predicate on integer states is primitive recursive.
There is a function in $\mathcal C_{\mathrm{Mono}}$ computed by no
three-coordinate Forward program under the fixed interface $I,O$.
\end{theorem}
\begin{proof}
The quotient has $M^n$ states. If its first $M^n$ observations contain
no stopping state, either they already contain a repetition or the
next observation repeats one of them. Determinism then makes all
subsequent observations cycle without stopping. This
proves~\eqref{eq:forward-bound}.
The finite iteration and guard are primitive recursive by
Lemma~\ref{lem:effective}, and the bound is a constant for the fixed
program, so bounded search proves primitive-recursive halting.

For strictness, enumerate the ordinary partial-recursive codes and
let $g(e)$ be the result of the code numbered $e$ on input zero.
This is partial recursive and has an undecidable domain by the
halting theorem. Theorem~\ref{thm:characterization} realizes $g$
in the full calculus. A Forward realization would decide its domain
by applying the halting test to the fixed primitive-recursive input
$I(e)$, a contradiction.
\end{proof}

The argument uses only a finite autonomous observation carrying the
stopping decision. In particular it gives a general principle beyond
this syntax.
\begin{corollary}[Closed observation barrier]\label{cor:barrier}
Suppose $T:S\to S$ has an observation $\pi:S\to F$ into a finite
set with a fixed finite encoding, $\pi T=\overline T\pi$, and
stopping is $\overline h\pi$.
If an input initialization $J$ has a computable observation $\pi J$,
then the halting domain on these inputs is decidable.
\end{corollary}
\begin{proof}
Start at $\pi J(n)$ and inspect the first $|F|$ finite observations.
The preceding repetition argument decides whether a stopping
observation occurs. The hidden state space $S$ may be infinite.
\end{proof}

\subsection{Descent and the depth of observation}

Descent breaks the closed residue argument in an exact way. The
following result concerns all integers, making its quantifiers
independent of any particular word encoding.

\begin{proposition}[Exact resolution of descent]\label{prop:resolution}
Let $B>0$, $m\geq2$, $N\in\Z$, and $r\in\N$. Then
\begin{equation}\label{eq:resolution}
 \left[\ \forall x,y\in\Z,\quad x\equiv y\pmod N
       \Longrightarrow
       x\div B^r\equiv y\div B^r\pmod m\ \right]
 \quad\Longleftrightarrow\quad B^r m\mid N.
\end{equation}
If $B\geq2$, preservation at every depth $r$ is equivalent to $N=0$,
where congruence modulo zero is equality.
\end{proposition}
\begin{proof}
First prove the one-step assertion for a positive divisor $A$.
If $Am\mid N$, replacing $x$ by $x+Nk$ changes its quotient by
$Nk/A$, a multiple of $m$. For necessity, write
$N=Aq+s$ with $0\leq s<A$. Comparing the congruent inputs $0,N$
gives $q\equiv0\pmod m$. If $s>0$, compare the congruent inputs
$A-s$ and $A-s+N$. Their quotients are $0$ and $q+1$, which
would give $q+1\equiv0\pmod m$, a contradiction. Thus $s=0$
and $Am\mid N$. Apply this with $A=B^r$; repeated Euclidean
division by $B$ equals division by $B^r$.
If every $B^rm$ divides a nonzero $N$, then $B^r\leq|N|$ for
all $r$, contradicting unbounded growth of the powers. Conversely,
congruent inputs modulo zero are equal and remain equal at every depth.
\end{proof}

For the cubic tower, every further digit removal multiplies the
required input modulus by three. Forward execution keeps all future
stopping observations inside one finite quotient. Descent permits
future queries to depend on successively deeper memory, and the full
language uses this access to implement the two stacks. The strict
separation is a property of these languages; particular programs
containing descent can still halt immediately, for example when their
guard is constantly true.

\section{Formalization in Lean 4}
\label{sec:formalization}

The development uses Lean~4~\cite{demouraUllrich2021} and
Mathlib~\cite{mathlib2020}. Its proof architecture connects three
layers: actual covering-space transport, exact analytic execution on
an invariant lattice, and effective discrete computation. The final
statements combine these layers for the same finite program.

\subsection{Geometric states and dependent fibers}

A geometric state has type $\{e:E\mid p(e)=x\}$, so its type records
the base point $x$. A path from $x$ to $y$ transports between these
two fiber types. In \code{Covering.lean}, transport is defined by
Mathlib's \code{IsCoveringMap.monodromy}, applied to the path's
endpoint-fixed homotopy class under an explicit covering-map hypothesis.
Lift existence and uniqueness identify this operation with an actual
continuous lifted curve.

The dependent fiber types make endpoint compatibility a mathematical
obligation. A supplied lift carries a continuity proof, its projection
equation, and its specified initial value. To identify the terminal
fiber element with the intended state, \code{Lattice.segment_transport}
proves equality of their underlying points and uses subtype
extensionality. Concatenation then yields the finite-run transport law.
Thus the geometric certificate concerns actual lifts with compatible
endpoints throughout the run.

\subsection{Analytic execution and the grid invariant}

The inductive expression type in \code{CoverExpressions.lean} has
integer and real interpretations. Structural induction proves
\code{evalReal_int} and joint real analyticity. The quotient and
remainder identities use Euclidean integer division, including negative
inputs; the factor $2\pi$ connects normalized integers to the actual
retained angular branches.

Native memory is a real vector together with a proof of membership in
the embedded integer grid. Grid preservation makes the analytic return
a map on this subtype. Its value is computed by the real expression
interpretation; the integer witness certifies legality. The guard's
analytic indicator is proved to agree with the Boolean decision on
this same grid. These correspondences give the exact optional step
and then \code{Program.evalReal_init}, the complete partial-execution
equality~\eqref{eq:native-eval}.

\subsection{Complete results and effective representations}

The Turing lower bound composes Mathlib's verified stack and tape
translations with the cubic compiler. Two additional bridges recover
the required complete result. First, equality of all track projections
determines the entire packed tape. Second, a fixed source code has
a finite transition-closed control support. Restricting labels to
this support produces a finite control subtype; each target label
carries the membership proof supplied by transition closure. Forgetting
these certificates preserves the complete configuration and execution.
The pointed enumeration keeps the initial label zero, which gives
the program-independent input map.

For the upper bound, primitive-recursive certificates cover signed
integer arithmetic, finite expressions, guards, and transitions.
Stopped iteration is then partial recursive. The state encoding
explicitly uses the encoding associated with its chosen
\code{Primcodable} finite-vector representation. The fixed input and
total output maps have separate primitive-recursive proofs. These
choices align the numerical interface with the execution semantics in
\code{Structure.continuation_characterization}.

\subsection{Source and verification}

The Lean development is in \code{src/Monodromy/}, with root import
\code{src/Monodromy.lean}. Toolchain and dependency versions are
recorded in \code{lean-toolchain}, \code{lakefile.lean}, and
\code{lake-manifest.json}.
Appendix~\ref{app:lean} gives the pinned versions and the correspondence
between the paper's statements and formal declarations. From the
repository root, verification uses
\begin{verbatim}
lake build
lake env lean -DautoImplicit=false -DwarningAsError=true src/Audit.lean
\end{verbatim}
The audit checks every exported \code{Monodromy} declaration and its
transitive proof dependencies against the standard Lean axioms
\code{propext}, \code{Quot.sound}, and \code{Classical.choice}.

\section{Discussion and Conclusion}
\label{sec:discussion}

\subsection{The computational role of deeper observation}

The construction organizes computation around three related structures.
An infinite covering fiber retains an unbounded discrete state. A
finite covering observation turns selected information into analytic
control. Descent changes which part of that retained state becomes
available to the same finite observation. Affine writing and finite
feedback complete the execution mechanism. The resulting three-coordinate
calculus has exactly classical partial-recursive power, with a geometric
realization of its finite runs.

The distinction between retention and access matters. An infinite
state space by itself can still support only decidable stopping when
a closed finite observation governs every future decision.
Corollary~\ref{cor:barrier} states this obstruction without assuming
a particular geometry. In the present calculus,
Proposition~\ref{prop:resolution} measures exactly how descent escapes
a fixed residue observation, and Theorem~\ref{thm:forward} converts
that distinction into a separation of computational power.

The covering groups needed here are elementary: the logarithmic
memory has integer translations, and the finite power observations
have cyclic monodromy. Conditional execution is supplied by
state-dependent analytic feedback. Thus this realization already
arises from abelian covering actions; its computational organization
lies in how retained branches, finite queries, and descent interact.
Transport composition and homotopy invariance govern the paths that
implement that organization.

\subsection{Physical realization and path deformation}

Homotopy invariance gives exact protection of branch transport against
endpoint-fixed path deformations within the covering base. A physical
realization would require a quantitative account of the perturbations
that preserve this class, together with reliable branch observation,
internal selection, and path generation. The robustness techniques of
\cite{gracaCampagnoloBuescu2005,gracaZhong2023} provide a starting point
for studying the analytic selectors and returns away from the exact
integer lattice.

Optical phase transport and fluid circulation suggest possible media
in which to investigate winding-based memory. A corresponding device
model would have to specify its physical state space and covering,
construct the read and write operations, and bound their precision,
energy, and execution costs. The discrete-return semantics identifies
the mathematical operations that such an implementation would need
to realize.

\subsection{Open problems}

A broader mathematical question is which effective continuation
systems admit an analogous combination of retained state, separating
finite observations, and recursive access to deeper structure.
The present finite calculus gives one complete instance and a general
finite-observation obstruction. It provides a precise starting point
for studying that question across other analytic coverings.

Within the calculus, further sublanguage classifications can ask which
restrictions admit a finite closed stopping observation and which
extensions escape it. Quantitative questions concern the size of
compiled query trees, the growth of memory codes, and the cost of
realizing a return. Reusable formal interfaces for covering transport,
analytic observations, and complete-execution refinement could support
these investigations across different geometric models.

The present model joins persistent branches, finite analytic control,
and access to deeper memory in one finite syntax. Its execution has
both an exact partial-recursive characterization and an actual
path-lifting realization. The Forward separation makes the contribution
of deeper access visible as a theorem about computational power.

\paragraph{AI contributions and author responsibility.}
Generative AI (OpenAI ChatGPT and Codex, including collaborating AI
agents) contributed substantially to the conception and mathematical
development of this work, including research ideation, theoretical
constructions, conjectures, proofs, Lean formalization, implementation,
and manuscript preparation.

The human author directed the research, determined its scope and
principles, established its aesthetic aims and the use of Lean as the
native research environment, critically evaluated the results, and
assumes full responsibility for the work.

The formalization and source code accompany this manuscript for
independent verification.

\clearpage
\appendix

\section{The fixed input and output maps}\label{app:io}

This appendix specifies the interface of~\eqref{eq:io} without
reference to the function being computed. It is inherited from the
fixed alphabet of Mathlib's partial-recursive compiler, and the
Monodromy block encoding is applied only after that alphabet is fixed.

Let $K=\{\mathsf{main},\mathsf{rev},\mathsf{aux},\mathsf{stack}\}$
be four track names and
$A=\{\mathsf{cons}_\ell,\mathsf{cons},\mathsf{bit}_0,
\mathsf{bit}_1\}$ four stack symbols. Use the finite tape alphabet
\[
 \Sigma=\{\mathsf{false},\mathsf{true}\}
       \times (K\to\operatorname{Option}A).
\]
The Boolean marks a stack bottom, and each track stores either a
symbol or an empty entry. The blank is the false marker with every
entry empty. Fix an enumeration $e:\Sigma\simeq[a]$ sending this
blank to zero. This same enumeration is used for every program,
input, and output. In Lean it is obtained from a fixed finite
enumeration by swapping the blank's index with zero.

For $s\in A$, let $\operatorname{col}(s)$ be the letter with bottom
marker false, symbol $s$ on the main track, and empty other tracks.
Let $\operatorname{mark}(s)$ be the same letter with marker true.
Write $\bin(n)$ for the ordinary binary digits of $n$ in
most-significant-first order, with $\bin(0)=[]$. The input word is
\begin{equation}\label{eq:numeric-word}
 W_n=\operatorname{mark}(\mathsf{cons})::
       \operatorname{map}(\operatorname{col},\bin(n)),
\end{equation}
interpreting each binary digit as its corresponding stack symbol.
This is exactly the compiler's packed word for the singleton numeric
list $[n]$: the stack's little-endian bits and final separator are
reversed when packed into the tape. Its first separator column carries
the bottom marker. The native input is
\begin{equation}\label{eq:input-exact}
 I(n)=\bigl(e(\operatorname{head}(W_n)),\ 0,\
              E_a(\operatorname{map}(e,\operatorname{tail}(W_n)))\bigr).
\end{equation}
The current symbol occupies the first coordinate because the initial
control label is zero in~\eqref{eq:configuration-code}.

Define a total block decoder $\operatorname{dec}_a:\N\to[a]^*$.
Starting from $(v,[])$, take exactly $v$ iterations of this rule:
look up the residue modulo $3^a$ in the finite list
$(\delta_j)_{j\in[a]}$; if it names $j$, replace the state by
$(v\div3^a,j::w)$; otherwise replace it by $(0,w)$.
Finally reverse the accumulated word. Zero has no matching letter,
so it remains stationary. The inequalities
$0<\delta_j<3^a$ give
\[
 \operatorname{dec}_a(E_a(w))=w,\qquad |w|\leq E_a(w).
\]
The second inequality proves that the chosen iteration budget always
suffices for valid encodings. The bounded definition gives a total
primitive-recursive map on every natural input.

For $x\in\Z^3$, form
$w=\operatorname{map}(e^{-1},\operatorname{dec}_a(\max(x_2,0)))$.
Fold this word from left to right, starting with zero. A main-track
$\mathsf{bit}_b$ replaces the accumulator $z$ by $2z+b$;
every other symbol leaves it unchanged. Call the resulting number
$O(x)$. The marker and the other three tracks are ignored.
Thus $O$ is total, independent of the source table, and reads the
canonical final right half-tape as its binary value. Since appended
blank symbols leave the fold unchanged, the readout is well-defined
on blank-extended half-tapes. The compiler leaves the separator in
the current cell and the numeric bits in the right half-tape at
termination, so reading only coordinate two recovers the entire
numeric answer.

All constants and finite tables here are fixed. In particular, the
input and decoder introduce no source-dependent preprocessing or
readout. The formal certificates of primitive recursiveness use a
fixed primitive coding of the finite alphabet and of integer vectors.

\section{Proof correspondence and reproducibility}\label{app:lean}

The project pins Lean to \code{leanprover/lean4:v4.35.0} and Mathlib
to commit
\begin{center}\small\code{56a005855ea17e0b0958f73ec801ae140b0a1679}.\end{center}
Definitions and proofs are in \code{src/Monodromy/}. The following
table lists the principal bridges; names are relative to the
\code{Monodromy} namespace.

\begingroup\small
\setlength{\tabcolsep}{4pt}
\renewcommand{\arraystretch}{1.05}
\begin{longtable}{@{}>{\raggedright\arraybackslash}p{0.31\textwidth}
  >{\raggedright\arraybackslash}p{0.65\textwidth}@{}}
\toprule
Mathematical statement & Formal entry points\\
\midrule
\endfirsthead
\toprule
Mathematical statement & Formal entry points\\
\midrule
\endhead
\bottomrule
\endfoot
Covering transport & \code{Transport.transport_trans},
 \code{Transport.transport_homotopic}\\
Logarithmic branches & \code{Logarithm.transport_turn}\\
Regular-root selectors & \code{Choice.kernel_fiber},
 \code{Choice.analyticAt_kernel}, \code{Choice.kernel_comp}\\
Lemma~\ref{lem:interpretation} &
 \code{CoverExpressions.evalReal_int},
 \code{CoverExpressions.analyticAt_evalReal},
 \code{CoverExpressions.Guard.evalReal_int},
 \code{CoverExpressions.Guard.eval_gate}\\
Complete native equality &
 \code{CoverExpressions.Program.evalReal_init}\\
Proposition~\ref{prop:lifting} &
 \code{CoverExpressions.Program.run_transport},
 \code{CoverExpressions.query_transport},
 \code{CoverExpressions.query_integer}\\
Finite ternary tables &
 \code{CoverExpressions.Ternary.eval_selectPower},
 \code{CoverExpressions.Ternary.eval_quotientPower}\\
Block word encoding & \code{BinaryBlocks.encode_cons},
 \code{BinaryBlocks.encode_injective},
 \code{BinaryBlocks.decode_encode}\\
Lemma~\ref{lem:one-step} & \code{CoverCompiler.stepInt_code},
 \code{CoverCompiler.code_injective}\\
Theorem~\ref{thm:universality} &
 \code{TwoStacks.decode_surjective}, \code{TwoStacks.eval_decode},
 \code{FiniteRenaming.eval_configuration},
 \code{CoverCompiler.evalInt_code},
 \code{CoverCompiler.advance_iterate_realize},
 \code{CoverCompiler.run_transport}; see also
 \code{Universality.every_finite_machine} for the direct radix realization\\
Classical compiler bridge & \code{FormalCompilation.every_partrec},
 \code{FormalCompilation.eval1}, \code{FormalCompilation.eval0},
 \code{NativeLowerBound.every_partrec}\\
Lemma~\ref{lem:io} & \code{NativeIO.primrec_input},
 \code{NativeIO.primrec_output}, \code{NativeIO.output_represents}\\
Lemma~\ref{lem:effective} & \code{IntegerPrimrec.add},
 \code{CoverExpressions.Program.primrec_stepInt},
 \code{CoverExpressions.Program.partrec_evalInt}\\
Theorem~\ref{thm:characterization} &
 \code{Structure.continuation_characterization},
 \code{Structure.analytic_continuation_realization},
 \code{Structure.computes_iff}\\
Lemma~\ref{lem:forward-residue} &
 \code{CoverExpressions.Program.forward_supported},
 \code{CoverExpressions.Program.forward_finite_observation}\\
Theorem~\ref{thm:forward} &
 \code{CoverExpressions.Program.forward_halting_bound},
 \code{ForwardBoundary.primrec_halting},
 \code{Structure.descent_strictly_extends}\\
Corollary~\ref{cor:barrier} &
 \code{Structure.closed_observation_barrier}\\
Proposition~\ref{prop:resolution} &
 \code{Resolution.iterate_tail_degree_iff},
 \code{Resolution.all_depths_iff}\\
\end{longtable}
\endgroup

The formal syntax allows any positive query degree; the predicate
\code{UsesRadix 3} selects exactly the cubic expressions of this
paper. Its constructors use parameter $2$ to specify degree $2+1=3$.
The formal definition \code{Structure.Computes} uses the result graph
of native \code{evalReal} directly; \code{computes_iff} proves its
equivalence with the complete partial equality~\eqref{eq:computes}.

The verification commands and axiom audit are given in
Section~\ref{sec:formalization}.
The main equivalence is an existence theorem for a finite realizing
program for each partial recursive function; its proof uses finite
enumerations and finite-support restrictions. This is the formal
compilation interface used in the mathematical results above.

\begingroup
\raggedright
\bibliographystyle{plainnat}
\bibliography{references}

\begin{thebibliography}{12}
\providecommand{\natexlab}[1]{#1}
\providecommand{\url}[1]{\texttt{#1}}
\expandafter\ifx\csname urlstyle\endcsname\relax
  \providecommand{\doi}[1]{doi: #1}\else
  \providecommand{\doi}{doi: \begingroup \urlstyle{rm}\Url}\fi

\bibitem[Blum et~al.(1989)Blum, Shub, and Smale]{blumShubSmale1989}
Lenore Blum, Mike Shub, and Steve Smale.
\newblock On a theory of computation and complexity over the real numbers:
  {NP}-completeness, recursive functions and universal machines.
\newblock \emph{Bulletin of the American Mathematical Society (New Series)},
  21\penalty0 (1):\penalty0 1--46, 1989.
\newblock \doi{10.1090/S0273-0979-1989-15750-9}.
\newblock URL
  \url{https://profshub.github.io/articles/1989-On_a_theory_of_computation_and_complexity_over_the_real_numbers.pdf}.

\bibitem[Carneiro(2019)]{carneiro2019}
Mario Carneiro.
\newblock Formalizing computability theory via partial recursive functions.
\newblock In John Harrison, John O'Leary, and Andrew Tolmach, editors,
  \emph{10th International Conference on Interactive Theorem Proving ({ITP}
  2019)}, volume 141 of \emph{Leibniz International Proceedings in
  Informatics}, pages 12:1--12:17. Schloss Dagstuhl -- Leibniz-Zentrum f{\"u}r
  Informatik, 2019.
\newblock \doi{10.4230/LIPIcs.ITP.2019.12}.
\newblock URL
  \url{https://drops.dagstuhl.de/entities/document/10.4230/LIPIcs.ITP.2019.12}.

\bibitem[Carneiro and {The mathlib Community}(2020)]{mathlibComputability}
Mario Carneiro and {The mathlib Community}.
\newblock Modelling partial recursive functions using {Turing} machines.
\newblock {mathlib4} computability source, 2020.
\newblock URL
  \url{https://github.com/leanprover-community/mathlib4/blob/56a005855ea17e0b0958f73ec801ae140b0a1679/Mathlib/Computability/TuringMachine/ToPartrec.lean}.
\newblock Pinned source and imported machine translations; revision specified
  in Appendix~B.

\bibitem[de~Moura and Ullrich(2021)]{demouraUllrich2021}
Leonardo de~Moura and Sebastian Ullrich.
\newblock The {Lean 4} theorem prover and programming language.
\newblock In Andr{\'e} Platzer and Geoff Sutcliffe, editors, \emph{Automated
  Deduction -- {CADE} 28}, volume 12699 of \emph{Lecture Notes in Computer
  Science}, pages 625--635. Springer, 2021.
\newblock \doi{10.1007/978-3-030-79876-5_37}.
\newblock URL \url{https://lean-lang.org/papers/lean4.pdf}.

\bibitem[Floater and Lyche(2007)]{floaterLyche2007}
Michael~S. Floater and Tom Lyche.
\newblock Two chain rules for divided differences and {Fa\`a di Bruno}'s
  formula.
\newblock \emph{Mathematics of Computation}, 76\penalty0 (258):\penalty0
  867--877, 2007.
\newblock \doi{10.1090/S0025-5718-06-01916-8}.

\bibitem[Forster(1981)]{forster1981}
Otto Forster.
\newblock \emph{Lectures on {Riemann} Surfaces}, volume~81 of \emph{Graduate
  Texts in Mathematics}.
\newblock Springer-Verlag, New York, 1981.
\newblock \doi{10.1007/978-1-4612-5961-9}.
\newblock URL \url{https://link.springer.com/book/10.1007/978-1-4612-5961-9}.
\newblock Translated by Bruce Gilligan.

\bibitem[Gra{\c{c}}a and Zhong(2023)]{gracaZhong2023}
Daniel~S. Gra{\c{c}}a and Ning Zhong.
\newblock Analytic one-dimensional maps and two-dimensional ordinary
  differential equations can robustly simulate {Turing} machines.
\newblock \emph{Computability}, 12\penalty0 (2):\penalty0 117--144, 2023.
\newblock \doi{10.3233/COM-210381}.
\newblock URL \url{https://arxiv.org/abs/2109.15073}.

\bibitem[Gra{\c{c}}a et~al.(2005)Gra{\c{c}}a, Campagnolo, and
  Buescu]{gracaCampagnoloBuescu2005}
Daniel~S. Gra{\c{c}}a, Manuel~L. Campagnolo, and Jorge Buescu.
\newblock Robust simulations of {Turing} machines with analytic maps and flows.
\newblock In S.~Barry Cooper, Benedikt L{\"o}we, and Leen Torenvliet, editors,
  \emph{New Computational Paradigms}, volume 3526 of \emph{Lecture Notes in
  Computer Science}, pages 169--179. Springer, 2005.
\newblock \doi{10.1007/11494645_21}.
\newblock URL
  \url{https://sqigmath.tecnico.ulisboa.pt/pub/GracaDS/05-GCB-stable.pdf}.

\bibitem[Hatcher(2002)]{hatcher2002}
Allen Hatcher.
\newblock \emph{Algebraic Topology}.
\newblock Cambridge University Press, 2002.
\newblock ISBN 978-0-521-79540-1.
\newblock URL \url{https://pi.math.cornell.edu/~hatcher/AT/ATpage.html}.

\bibitem[Koiran and Moore(1999)]{koiranMoore1999}
Pascal Koiran and Cristopher Moore.
\newblock Closed-form analytic maps in one and two dimensions can simulate
  universal {Turing} machines.
\newblock \emph{Theoretical Computer Science}, 210\penalty0 (1):\penalty0
  217--223, 1999.
\newblock \doi{10.1016/S0304-3975(98)00117-0}.
\newblock URL \url{https://sites.santafe.edu/~moore/pubs/elem.html}.

\bibitem[Shannon(1941)]{shannon1941}
Claude~E. Shannon.
\newblock Mathematical theory of the differential analyzer.
\newblock \emph{Journal of Mathematics and Physics}, 20\penalty0
  (1--4):\penalty0 337--354, 1941.
\newblock \doi{10.1002/sapm1941201337}.
\newblock URL \url{https://onlinelibrary.wiley.com/doi/10.1002/sapm1941201337}.

\bibitem[{The mathlib Community}(2020)]{mathlib2020}
{The mathlib Community}.
\newblock The {Lean} mathematical library.
\newblock In \emph{Proceedings of the 9th {ACM SIGPLAN} International
  Conference on Certified Programs and Proofs}, pages 367--381. Association for
  Computing Machinery, 2020.
\newblock \doi{10.1145/3372885.3373824}.
\newblock URL
  \url{https://leanprover-community.github.io/papers/mathlib-paper.pdf}.

\end{thebibliography}
\endgroup
\end{document}